\documentclass[12pt,preprint,aps,showpacs,showkeys,floatfix]{revtex4}
\usepackage{bm}
\usepackage{amssymb}
\usepackage{amsmath}
\usepackage{amsfonts}
\usepackage{graphicx}
\usepackage{graphics}
\usepackage{float}

\begin{document}

\title{Maxwell-J{\bf $\ddot{\bf u}$}ttner distributions in relativistic molecular dynamics}

\author{A.Aliano} \email{antonio.aliano@polito.it} \affiliation{Dipartimento di
Fisica Politecnico di Torino}
\author{L.Rondoni}
\email{lamberto.rondoni@polito.it} \affiliation{Dipartimento di
Matematica Politecnico di Torino}
\author{G. P. Morriss}
\email{G.Morriss@unsw.edu.au} \affiliation{School of physics UNSW
Sydney}
%\date{today}
\begin{abstract}
In relativistic kinetic theory, which underlies relativistic
hydrodynamics, the molecular chaos hypothesis stands at the basis of
the equilibrium Maxwell-J$\ddot{\mbox{u}}$ttner probability
distribution for the four-momentum $p^{\alpha}$. We investigate the
possibility of validating this hypothesis by means of microscopic
relativistic dynamics. We do this by introducing a model of
relativistic colliding particles, and studying its dynamics. We
verify the validity of the molecular chaos hypothesis, and of the
Maxwell-J$\ddot{\mbox{u}}$ttner distributions for our model. Two
linear relations between temperature and average kinetic energy are
obtained in classical and ultrarelativistic regimes.
\end{abstract}



\maketitle
\section{introduction}
The study of relativistic fluids, both from the hydrodynamic and
kinetic point of view has been widely investigated
\cite{degr,ce,stew,uz,krem}. In this context, the relativistic
Boltzmann equation
\begin{equation}
p^{\,\nu}\frac{\partial f}{\partial x^{\nu}} + m_0 \frac{\partial f
F^{\nu}}{\partial p^{\,\nu}} = \int (f'_*f' - f_*f) \Omega
\frac{d^3{p'}}{{p'}^0}\frac{d^3p_*}{p^0_*}\frac{d^3p_*'}{{p'}^0_*}
\label{1} \ \ ,
\end{equation}
represents the best known tool, which is based on a molecular chaos
hypothesis, like the Boltzmann equation in classical kinetic theory.
Here $x^{\nu}$, $p^{\nu}$, $F^{\nu}$ are respectively the position,
momentum and force four-vectors, $m_0$ is the rest mass, $\Omega$ is
the interaction cross-section, and $f$ is the single particle
distribution function. Collisionless relativistic plasmas are
investigated by means of the relativistic Vlasov equation, obtained
neglecting the collision term in Eq.(\ref{1}). The equations of
relativistic Hydrodynamics, which macroscopically describe
relativistic fluids,
are derived also from Eq.(\ref{1}), similarly to the classical case.\\
The chaotic hypothesis, which underlies Eq.(\ref{1}), explains how
the microscopic components of a fluid reach a local equilibrium
state. Classically, it is well established that this is a
consequence of the interactions among the particles, as illustrated,
for instance, by molecular dynamics \cite{aw}.\\
In order to investigate the validity of the molecular chaos
assumption in relativistic kinetic theory, we propose a simple model
of $N$ relativistic colliding particles, and investigate the
properties of its dynamics.\\
In fact, to the best of our knowledge, many particle relativistic
systems have only been studied either from a kinetic or hydrodynamic
point of view, because the microscopic dynamics of such particle
systems presents many difficulties. For instance, it is highly
problematic to write covariant hamiltonians (and the related
4-vector equations of motion) for the systems. Other difficulties
concern: the choice of the reference frame, since every particle has
a different proper time; the form of the interaction potential,
since the action and reaction principle holds only for contact
interactions; the effects of length contraction and time dilation.
The consequence of this is that, as far as we know, no direct
microscopic evidence for the
molecular chaos hypothesis in relativistic dynamics has been provided. \\
To overcome this difficulty, we propose a non-covariant hamiltonian
written with respect to the center of mass frame, taken as the
Lorentz rest frame, which yields the non-covariant equations of
motion
\begin{eqnarray}
\left\{ \begin{array}{ll}
& \dfrac{d\mbox{\boldmath $x$}_j}{dt} = \dfrac{c\,\mbox{\boldmath $p$}_j}{\sqrt{\mbox{\boldmath $p$}_j^2 + m_0^2c^2}} \\
&  \\
& \dfrac{d\mbox{\boldmath $p$}_j}{dt} = \mbox{\boldmath
$F$}_{j}^{WCA}(\mbox{\boldmath $x$}) \,\,\,\,\,\,\,\,\,\,\,\,\,\
j=1,2,...,N \ \ ,
\end{array} \right. \label{me1}
\end{eqnarray}
where $N$ is the number of particles. For the force $\mbox{\boldmath
$F$}_{j}^{WCA}$ we propose to use
\begin{eqnarray}
\mbox{\boldmath $F$}_{j}^{WCA} = - \sum_{i\neq j}
\frac{\mbox{\boldmath $r$}_{ij}}{r_{ij}}\frac{\partial
\Phi^{WCA}_{ij}}{\partial r_{ij}} = \left\{ \begin{array}{ll}  &
\sum_{i\neq j}\dfrac{24\epsilon}{\sigma}\dfrac{\mbox{\boldmath
$r$}_{ij}}{r_{ij}}\left[2\left(\dfrac{\sigma}{r_{ij}}\right)^{13} -
\left(\dfrac{\sigma}{r_{ij}}\right)^7 \right] ;\,\,\
r_{ij}<2^{1/6}\sigma
\\ & 0 \,\,\,\,\,\,\,\,\,\,\,\,\,\,\,\,\,\,\,\,\,\,\,\,\,\,\,\,\,\,\,\,\,\,\,\,\,\,\,\,\,\,\,\,\,\,\,\,\,\,\,\,\,\,\,\,\,\,\,\,\,\,\,\,\,\,\,\,\,\,\,\,\,\,\,\,\,\,\,\,\,\,\ ;\,\,\
r_{ij}\geq 2^{1/6}\sigma
\end{array} \right. \label{me2}
\end{eqnarray}
where $\mbox{\boldmath $r$}_{ij} = (\mbox{\boldmath
$r$}_{i}-\mbox{\boldmath $r$}_{j})$, $r_{ij}=|\mbox{\boldmath
$r$}_{ij}|$, $\Phi^{WCA}_{ij}$ is the Weeks-Chandler-Andersen
interaction potential \cite{evmor}; the quantities $\epsilon$ and
$\sigma$ are obtained from the Lennard-Jones (LJ) potential which
defines $\Phi^{WCA}_{ij}$, and  represent respectively the depth of
the LJ potential, and the distance at which it changes sign.\\
Therefore, particles move according to the relativistic dynamics
when they do not interact, while their interactions are modelled
classically, so that the total momentum and the total kinetic energy
of particles are preserved by the collision process. Although this
is not completely rigorous, our  procedure meets all the microscopic
requirements of relativistic kinetic theory, i.e. the invariance of the momentum 4-vectors.\\
In this paper we simulate a 2D system of $N$ relativistic particles
(with $N=28$), through a MD algorithm, which implements the
equations of motion (\ref{me1},\ref{me2}) with periodic boundary
conditions, for a density $\rho=N/A=0.2$ (with $A$ the cell area),
which is not a low density case. The simulations are performed for
different initial kinetic energies corresponding to classical,
relativistic and ultrarelativistic regimes. Furthermore, we take
$\epsilon=\sigma=1$.\\
In the low density limit, the contribution of the collisions is
expected to become negligible, and the dynamics to tend to a fully
covariant dynamics.

\section{Results}

Our results show that the simulated systems all reach an equilibrium
state since their observables, such as the pressure, converge to an
equilibrium value, while the probability distribution functions
(PDFs) of the values of microscopic quantities like momentum $p_x$
and kinetic energy $\xi$ reach an invariant form. In particular, we
find that the PDFs of $p_x$ reduce to the Maxwell-Boltzmann (MB)
distribution in the classical limit, as desired. This is due to
chaos in the dynamics, which is evidenced by the fact that the
numerically evaluated largest Lyapounov exponents are positive.

\subsection{Probability Distribution Functions}

The standard relativistic kinetic theory predicts that the PDF of
$p^{\alpha}$ has the form of the Maxwell-J$\ddot{\mbox{u}}$ttner
(MJ) distribution, $f_{MJ}=d\exp{(-U^{\alpha}p_{\alpha}/k_BT)}$,
with $d$ a normalization constant and $U^{\alpha}$ the hydrodynamic
four-velocity (with $U^z=0$) \cite{degr,ce}. In the local rest
frame, $f_{MJ}$ can be written as
\begin{equation}
f_{MJ}(p_x,p_y) = d\,
\exp{\left(-a\sqrt{\frac{p_x^2+p_y^2}{m_0^2c^2}+1}\right)} \ \ ,
\label{mj}
\end{equation}
where $p_x$, $p_y$ are the spatial components of $p_{\alpha}$, $c$
is the speed of light, and where $d$ and $a$ are two constants
related by the normalization condition
\begin{equation}
d=\left(\frac{1}{2\pi m_0^2c^2}\right)\frac{a^2 e^a}{1+a} \ \ .
\label{norm}
\end{equation}
As well known \cite{degr,ce}, $a$ involves the temperature of the
system \footnote{ By definition \cite{mueller}, classical is the
regime with $a \gg 1 $ and ultrarelativistic the regime with $a \ll
1$.}, because
\begin{equation}
a=m_0c^2/k_BT \ \ . \label{T}
\end{equation}
Integrating Eq.($\ref{mj}$) over $p_y$, one obtains the PDF for
$p_x$ only:
\begin{eqnarray}
g_{MJ}(p_x) = 2 m_0c d \, \sqrt{\frac{p_x^2}{m_0^2c^2}+1}\,\,\cdot
K_1\left(a\sqrt{\frac{p_x^2}{m_0^2c^2}+1}\right) \ \ ,\label{mjpx}
\end{eqnarray}
where $K_1(x)$ is the modified K-Bessel function of first order.
Considering the kinetic energy $\xi=c\sqrt{\mbox{\boldmath $p$}^2
+m_0^2c^2}-m_0c^2$, Eq.($\ref{mj}$) can also be rewritten as
\begin{eqnarray}
h_{MJ}(\xi)= \frac{2\pi}{c^2} d  \,(\xi
+m_0c^2)\cdot\exp{\left(-a\frac{\xi+m_0c^2}{m_0c^2}\right)} \ \
.\label{mje}
\end{eqnarray}
It is interesting to observe that, if an expression like the MB
distribution was written for the relativistic $p_x$, i.e. if one
started from
\begin{equation}
f_{MB}(p_x,p_y) = f_{MB}(p_x)f_{MB}(p_y) = \frac{\tilde{a}}{\pi
m_0^2c^2}\,
\exp{\left(-\tilde{a}{\frac{p_x^2+p_y^2}{m^2_0c^2}}\right)} \ \
,\label{mb}
\end{equation}
the PDF of the relativistic kinetic energy $\xi$, after some
calculations, would take the form
\begin{equation}
h_{MB}(\xi) =
\frac{2\tilde{a}e^{\tilde{a}}}{m_0^2c^4}\,(\xi+m_0c^2)\,\exp{\left(-\tilde{a}\frac{(\xi+m_0c^2)^2}{m_0^2c^4}\right)}
\ \ .\label{mbe}
\end{equation}
Comparing Eq.s ($\ref{mb},\ref{mbe}$) with Eq.s
($\ref{mjpx},\ref{mje}$), one notices that the MJ distribution is
not merely the MB distribution with the relativistic
$p_x$ and $\xi$ in place of the classical momentum and kinetic energy.\\
We fit the histograms constructed through our MD simulations to the
PDFs given above, and for simplicity we take $m_0=c=1$.\\
The following figures are obtained for different mean kinetic
energies, where the mean kinetic energy is the time average of the
total kinetic energy divided by the number of particles. The
histograms are constructed recording the instantaneous values of
momentum $p_x$ and kinetic energy $\xi$ for a given particle. This
operation is repeated every 200 timesteps, in order to decorrelate
the recorded data.
\begin{figure}[H]
\centerline{\framebox{\bf Mean kinetic energy per particle =
$9.87\times 10^{-2}$}}
       \includegraphics[width=8cm,height=8cm]{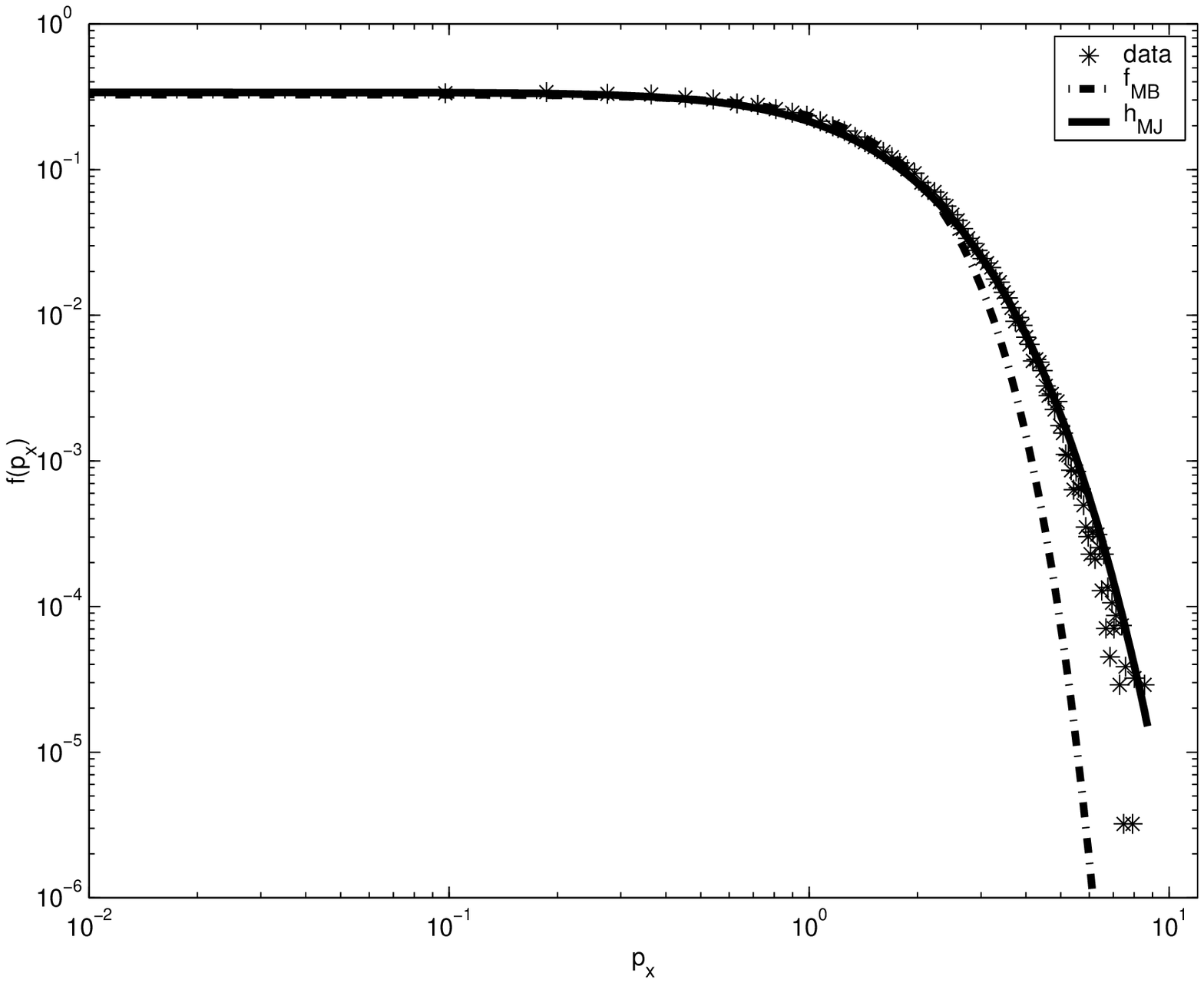}
       \includegraphics[width=8cm,height=8cm]{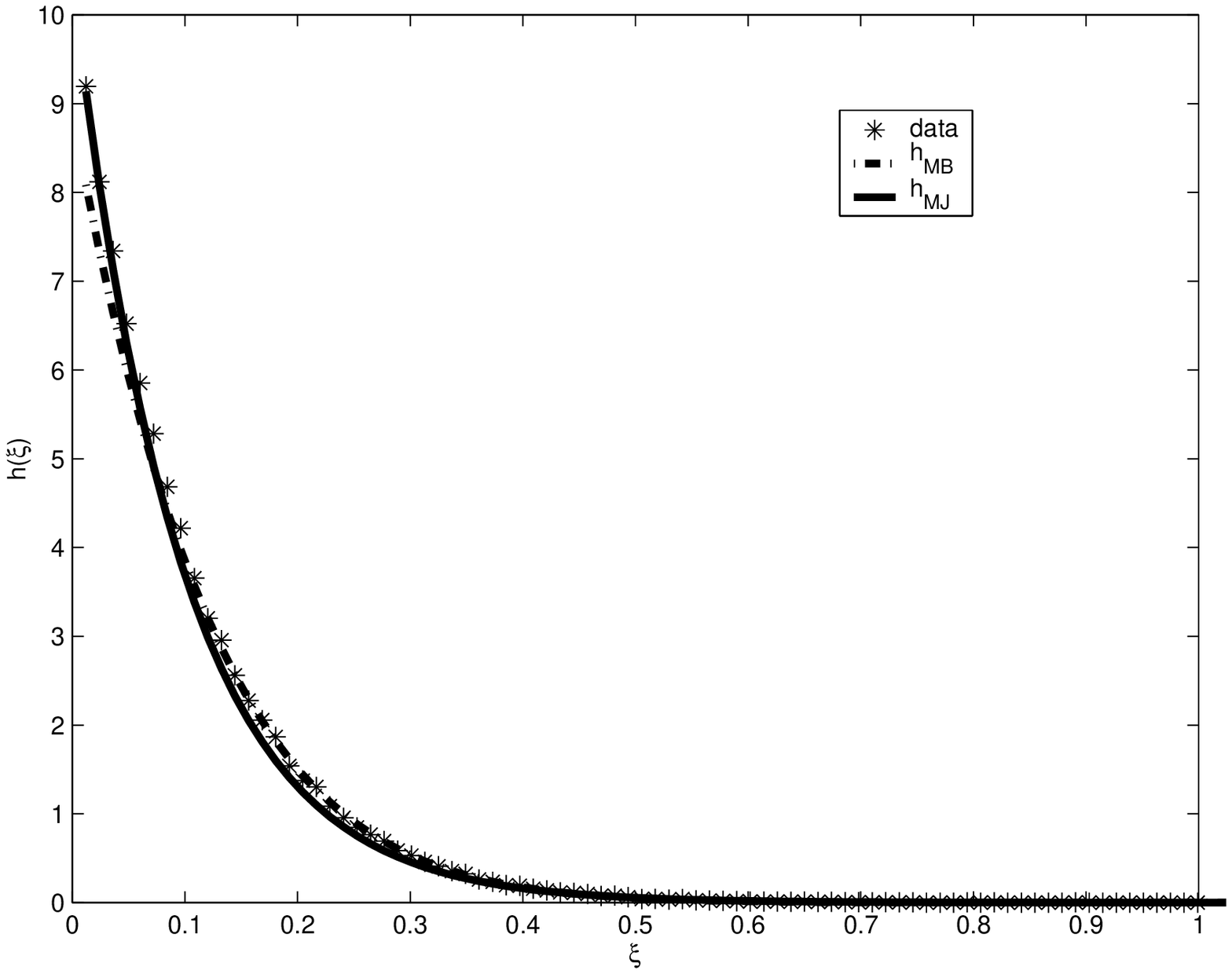}
       \caption{\label{1} \scriptsize{Fit of the data for momentum $p_x$ on a log scale (left panel).
       Fitted histograms of the kinetic energy $\xi$ in linear scale
       (right panel). The parameter of the Maxwell-J$\ddot{\mbox{u}}$ttner PDFs takes the value $a=10.4875$ for the PDF of$p_x$ yielding
       $k_BT=0.095$ (left panel), and $a=11.2595$ for the PDF of $\xi$ leading
       to $k_BT=0.089$ (right panel). The classical Maxwell Boltzmann PDFs fits well the
       data only in the low energy cases.}}
\end{figure}
\begin{figure}[H]
\centerline{\framebox{\bf Mean kinetic energy per particle = $9.83
\times 10^{-1}$ }}
       \includegraphics[width=8cm,height=8cm]{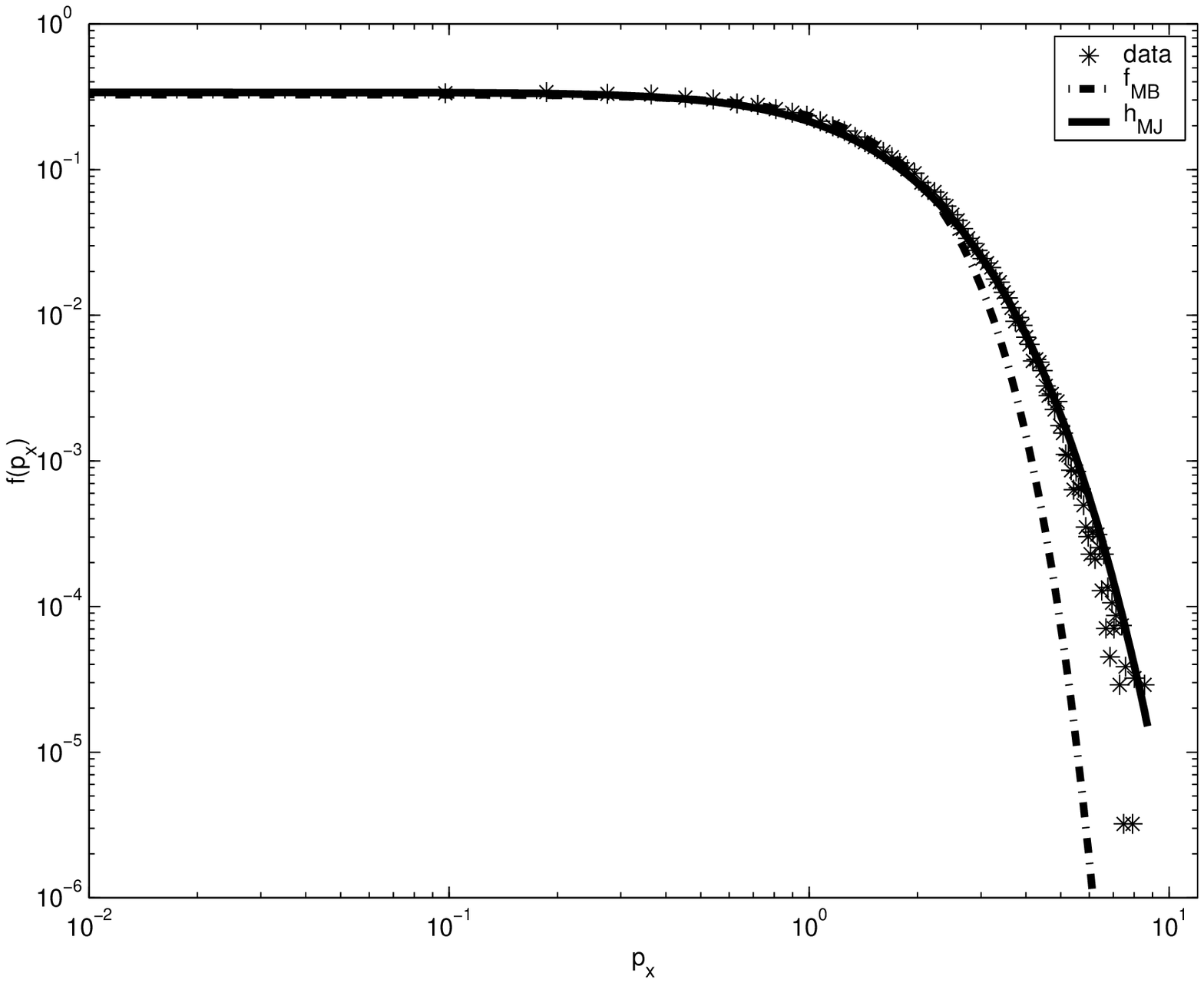}
       \includegraphics[width=8.cm,height=8.cm]{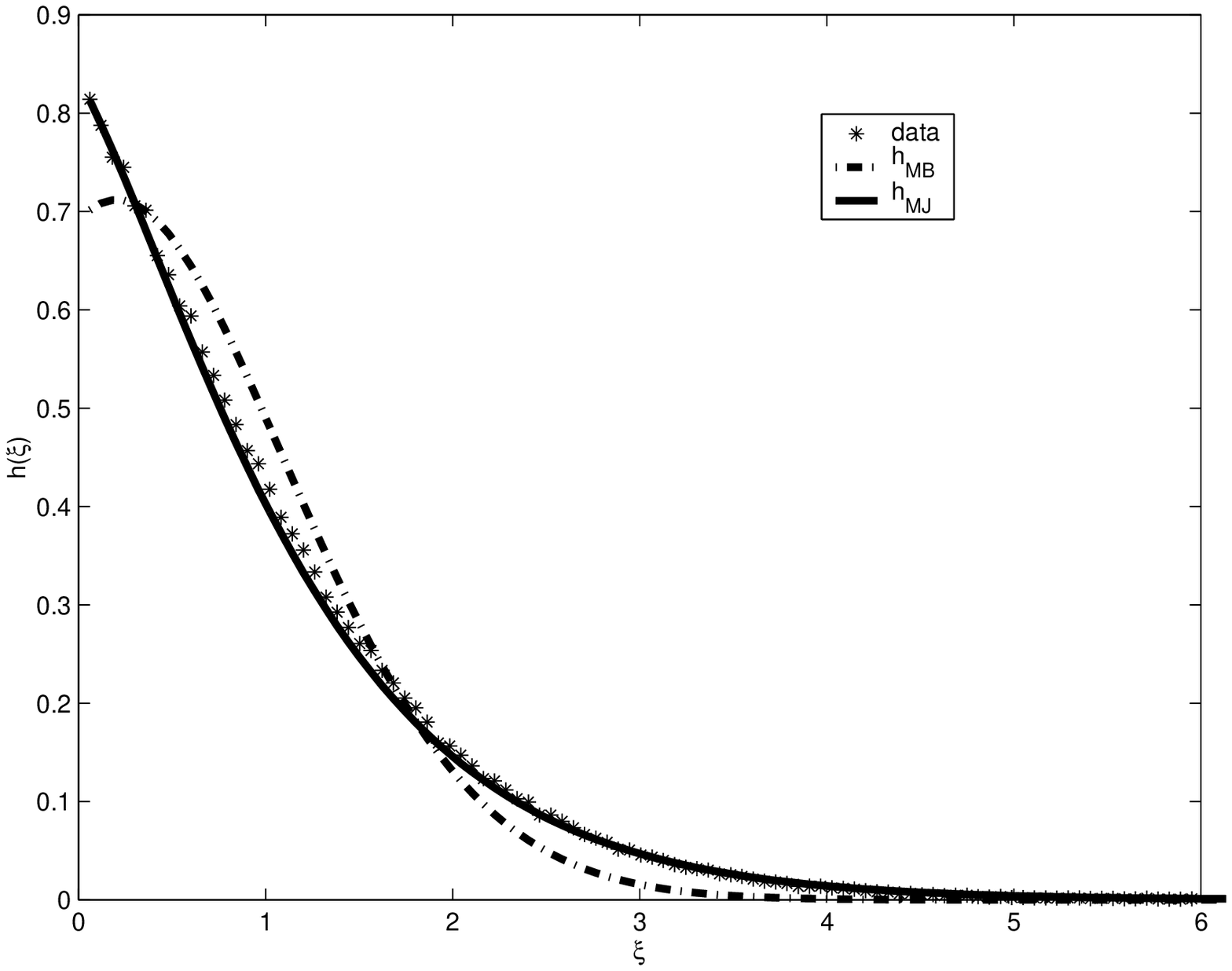}
       \caption{\label{2} \scriptsize{Fit of the data for momentum $p_x$ on a log scale (left panel).
       Fitted histograms of the kinetic energy $\xi$ in linear scale
       (right panel). The parameter of the Maxwell-J$\ddot{\mbox{u}}$ttner PDFs takes the value $a=1.4067$ for the PDF of $p_x$ yielding
       $k_BT=0.711$ (left panel), and $a=1.4225$ for the PDF of $\xi$ leading
       to $k_BT=0.703$ (right panel). The classical MJ distributions fits
       better the data than the MB ones at these energies.}}
\end{figure}
\begin{figure}[H]
\centerline {\framebox{\bf Mean kinetic energy per particle =
$6.87$}}
       \includegraphics[width=8cm,height=8cm]{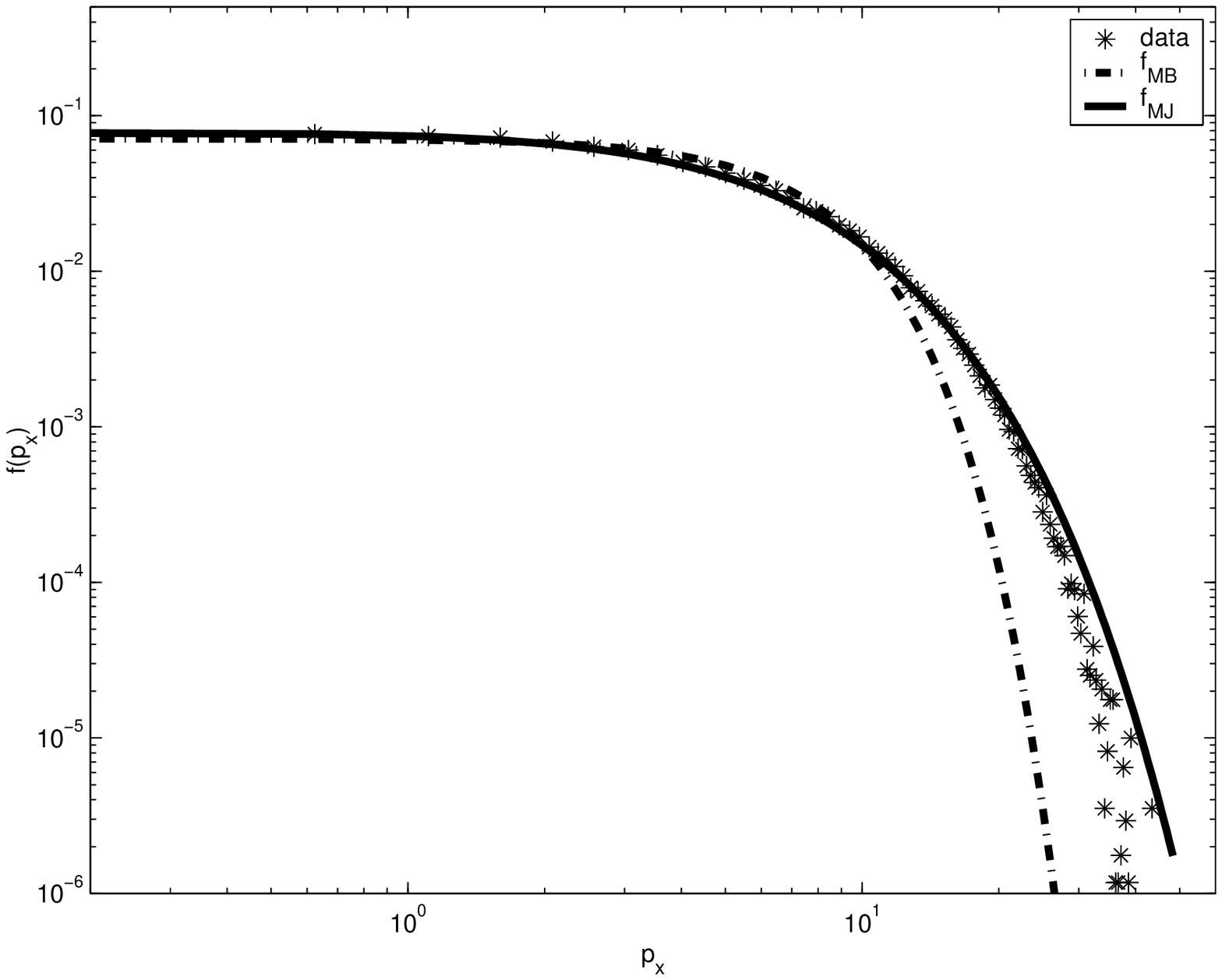}
       \includegraphics[width=8.cm,height=8.cm]{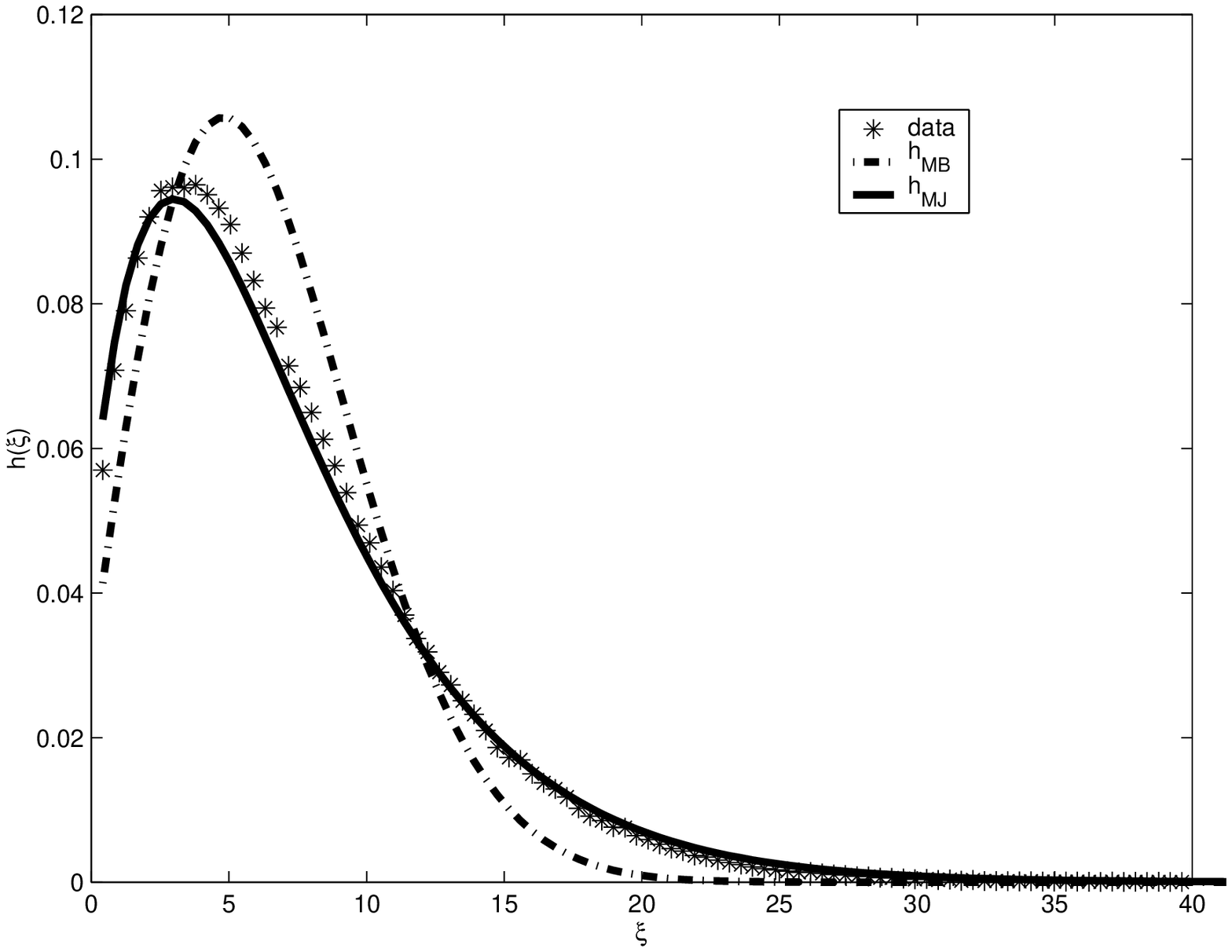}
       \caption{\label{3} \scriptsize{Fit of the data for momentum $p_x$ on a log scale (left panel).
       Fitted histograms of the kinetic energy $\xi$ in linear scale
       (right panel). The parameter of the Maxwell-J$\ddot{\mbox{u}}$ttner PDFs takes the value $a=0.2540$ for the PDF of $p_x$ yielding
       $k_BT=3.937$ (left panel), and $a=0.2501$ for the PDF of $\xi$ leading
       to $k_BT=3.998$ (right panel). The MB does not fit the data at these ultrarelativistic energies.}}
\end{figure}
For the Maxwell-J$\ddot{\mbox{u}}$ttner PDFs, if the parameters $a$
and $d$ are obtained as independent parameters by fitting the
numerical data to Eq.s (\ref{mjpx},\ref{mje}), the normalization
condition (\ref{norm}) both for $p_x$ and $\xi$ is verified. This
indicates that the MJ-PDF is indeed appropriate for our data, and
that the data are consistent.

\subsection{Measurement of temperature for a relativistic system}

The microscopic definition of the temperature of a system composed
by relativistic particles is an open issue \cite{lands}. However,
the Maxwell-J$\ddot{\mbox{u}}$ttner PDF contains one parameter,
which, in analogy with the classical Maxwell-Boltzmann PDF is
identified with the quantity $k_BT$. Therefore, observing that the
Maxwell-J$\ddot{\mbox{u}}$ttner PDFs fit well our histograms, it
becomes reasonable to assume for our system $a^{-1}=k_BT$ as a
definition of temperature  obtained from the microscopic dynamics.
\begin{figure}[H]
\centerline{\framebox{\bf Temperature vs Mean kinetic energy per
particle}} \centering
       \includegraphics[width=8.cm,height=8.cm]{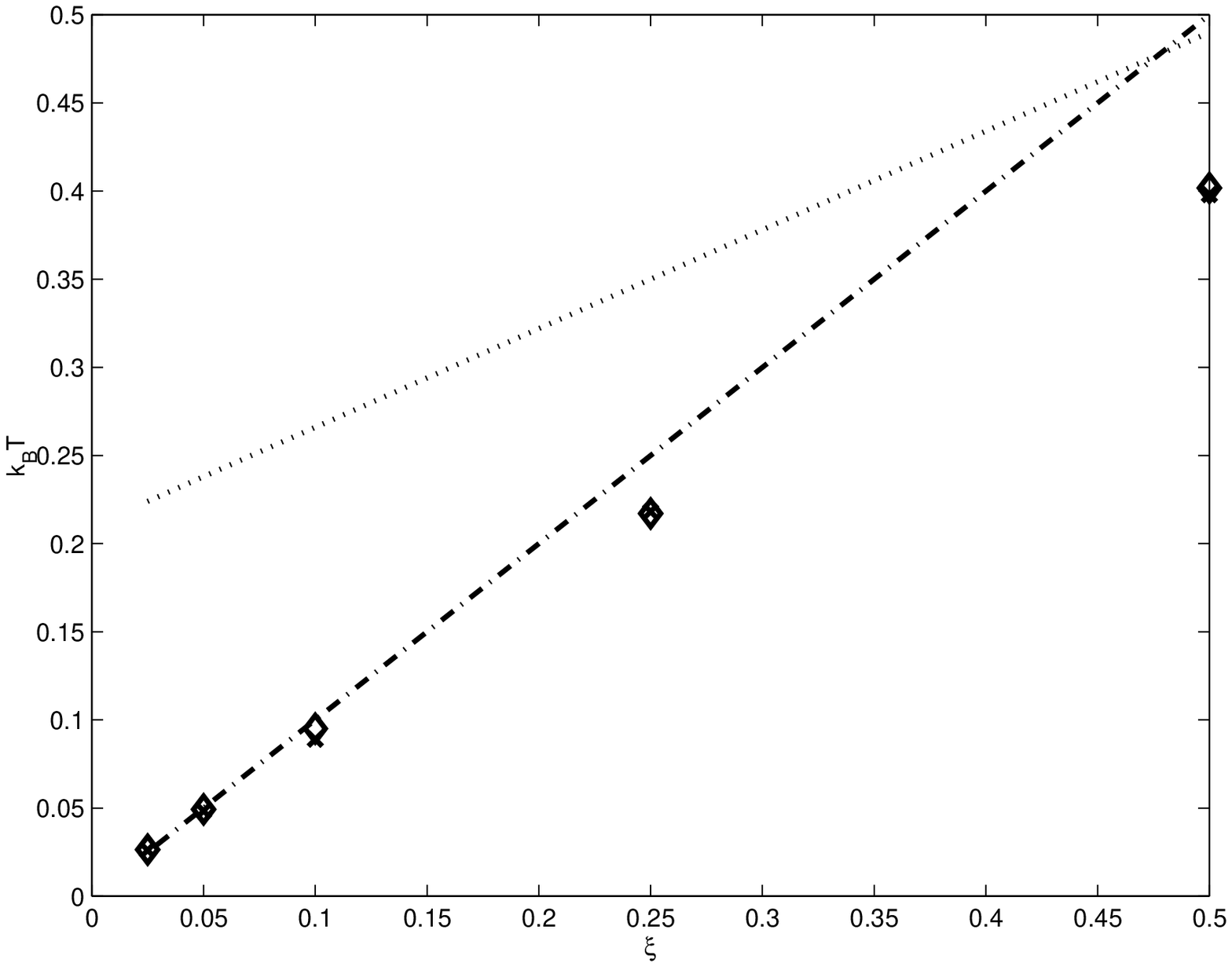}
       \includegraphics[width=8.cm,height=8.cm]{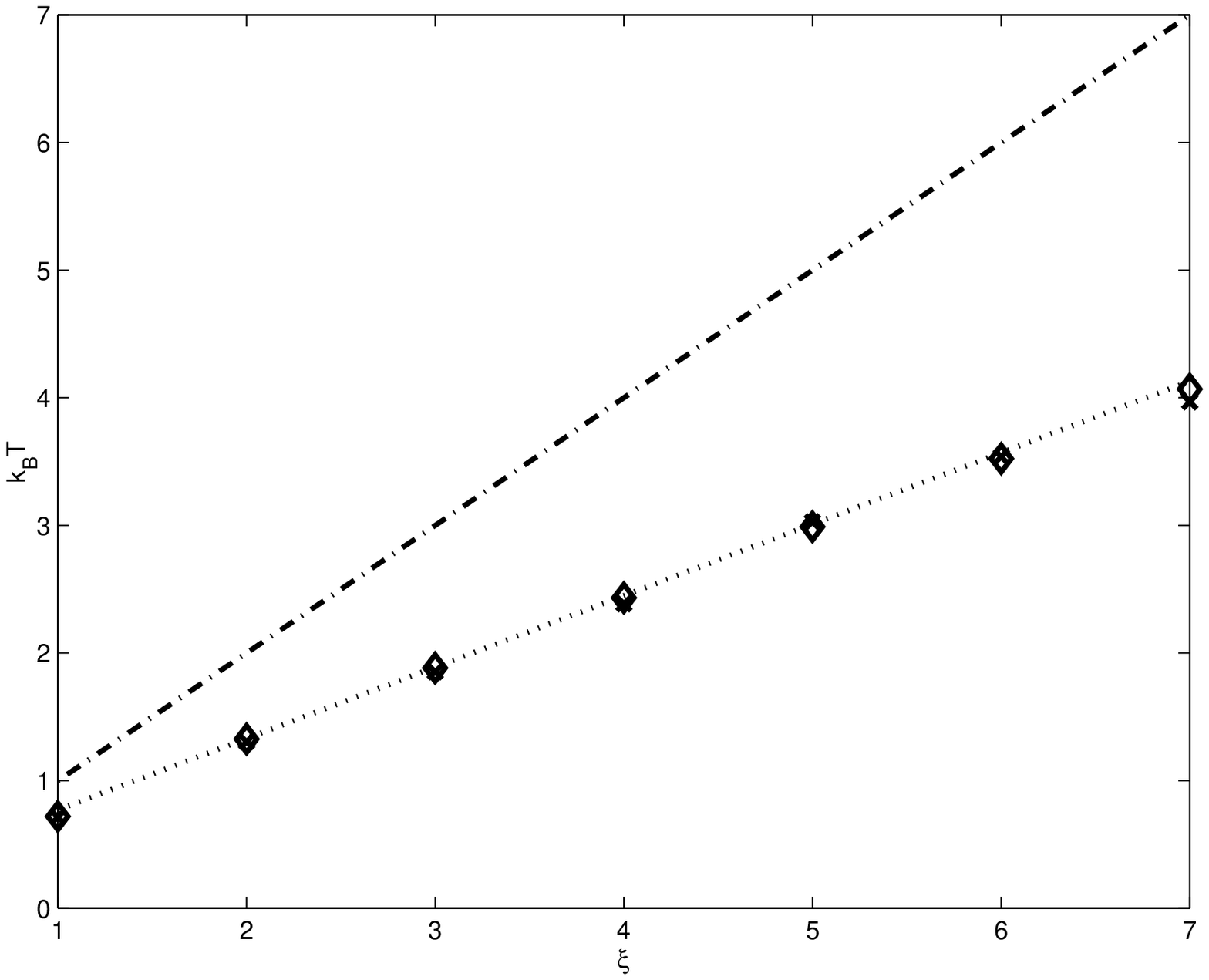}
       \caption{\label{4} \scriptsize{Plot of temperature vs mean kinetic energy per particle.
       Temperatures are calculated through Eq.($\ref{T}$) from the fitting parameter $a^{-1}=k_BT$, obtained both for
       the kinetic energy($*$) and the $p_x$ ($\diamond$).
       The two different calculations yield indistinguishable values in this figure.
       In the left panel the classical regime is plotted, while in the right one the
       ultrarelativistic limit is shown. In both panels the dotted lines represent
       the relation between $k_B$T and $\xi$ valid in the low energy cases; the dash-dotted lines
       the relation between $k_B$T and $\xi$ valid in the high energy cases.}}
\end{figure}
A linear relation between this temperature and the mean value of the
kinetic energy per particle has been found for the classical and
ultrarelativistic cases. For $k_BT=a^{-1}\lesssim 0.1$ (classical
regime), the relation was found to be, as expected, $k_BT=\xi$,
while for $k_BT=a^{-1}\gtrsim 1$ (relativistic and ultrarelativistic
regimes), we verified a linear relation of the form
$k_BT=0.56\xi+0.21$. The transition between the two regimes takes
place in a small range of kinetic energy values.

\section{Conclusions}

In this paper we have tested a 2D molecular dynamics model intended
to simulate the microscopic dynamics of $N$ relativistic colliding
particles, with total constant energy $E$, and have observed its
relaxation to an equilibrium state.\\ Our model satisfies the
requirements of momentum and kinetic energy conservation before and
after the collisions, underlying the equilibrium relativistic
kinetic theory. The histograms found by these simulations for the
momentum $p_x$, and for the kinetic energy $\xi$ are well fitted by
the PDFs of the standard relativistic kinetic theory, i.e. by the
PDFs derived from the MJ distributions. In addition to this, the
statistics of the dynamics of our model reduces to the classical one
when the kinetic energy takes small
values.\\
Our model suffers from the difficulties of not being fully
relativistic, because the particle interactions are treated
classically; therefore, it becomes more and more acceptable as the
particle density decreases, or the collision rate tends to zero
making the dynamics tend to a fully covariant form. Moreover, as we
are going to report in \cite{alron}, reducing densities does not
produce any qualitatively different result, which indicates that in
the limit of low collision rates the macroscopic behaviour of our
systems is not substantially different from that of the higher
density cases. This, together with the observed validity of the MJ
distributions, provides a justification for our model, as a tool to
simulate relativistic many particle systems. Otherwise, if this
model is accepted, it affords a microscopic justification of the
relativistic molecular chaos hypothesis, underlying relativistic
kinetic theory
and relativistic hydrodynamics.\\
Furthermore, linear relations of temperature and mean kinetic energy
have been found both in classical and ultrarelativistic regimes.
This allows us to obtain a definition of temperature in a
relativistic system, something rather problematic in general
\cite{lands}, which deserves further investigations.

\section*{Acknowledgements}

The authors are grateful to Fasma Diele for help with data handling.

\end{document}